\documentclass[11pt, onecolumn]{IEEEtran}

\usepackage{booktabs,pifont, subcaption}
\usepackage[numbers,compress]{natbib}
\usepackage{amsmath,amssymb,amsthm,amsfonts,mathtools}
\usepackage{graphics,graphicx,color}
\usepackage{caption}
\usepackage{flushend}
\usepackage[doublespacing]{setspace}

\newcommand{\cmark}{\ding{51}}%
\newcommand{\xmark}{\ding{55}}%
\newcommand{\ma}[1]{\mathbf{#1}}

\newcommand{\argmin}[2]{\underset{#1}{\mathrm{argmin}}~{#2}}

\newcommand{\Rco}{\mathbf{R}_{\text{co}}}

\newcommand{\mPhi}[1]{\Phi(\mathbf #1)}
\newcommand{\mPsi}[1]{\Psi(\mathbf #1)}
\newcommand{\mOmega}[1]{\Omega(\mathbf #1)}
\newcommand{\mlambda}[1]{\boldsymbol{\lambda_{_{#1}}}}
\newcommand{\mlambdabar}[1]{\overline{\boldsymbol{\lambda}}_{_{#1}}}
\newcommand{\eqdef}{\triangleq}

\newtheoremstyle{mynewtheorem}
{2pt}
{2pt}
{\it}
{3pt}
{\bf}
{.}
{3pt}
{\thmname{#1}~\thmnumber{#2}}%

\theoremstyle{mynewtheorem}
\newtheorem{proposition}{Proposition }

\newtheorem{rem}{Remark}
\newtheorem{mdef}{Definition}

\begin{document}
\title{Structured Autocorrelation Matrix Estimation for Coprime Arrays}
\author{\thanks{ \protect\doublespacing This research is supported in part by the U.S. National Science Foundation, under grant OAC-1808582, and by the U.S. Air Force Office of Scientific Research, under the Young Investigator Program.

This is a preprint of work that has been submitted for publication to a peer-reviewed journal.}
\IEEEauthorblockN{Dimitris G. Chachlakis and Panos P. Markopoulos$^*$\thanks{$^*$Corresponding author.}}\\[.6cm]
\IEEEauthorblockA{Department of Electrical and Microelectronic Engineering\\ Rochester Institute of Technology\\ Rochester, NY 14623\\E-mail: \texttt{dimitris@mail.rit.edu} and \texttt{panos@rit.edu}}
}
\maketitle
\thispagestyle{empty}
\vspace{-1.0cm}
\begin{abstract}
A coprime array receiver processes a collection of received-signal snapshots to estimate the autocorrelation matrix of a larger (virtual) uniform linear array, known as coarray. By the received-signal model, this matrix has to be (i) Positive-Definite, (ii) Hermitian, (iii) Toeplitz, and (iv) its noise-subspace eigenvalues have to be equal. Existing coarray autocorrelation matrix estimates satisfy a subset of the above conditions. In this work, we propose an optimization framework which offers a novel estimate satisfying all four conditions. Numerical studies illustrate that the proposed estimate outperforms standard counterparts, both in autocorrelation matrix estimation error and Direction-of-Arrival estimation. 	
\end{abstract}
	
{\it{\textbf{Index Terms --}}} Sensor array processing, Coprime arrays, Coarray, Autocorrelation estimation.

\markboth{\it PREPRINT}{\it  PREPRINT}
	
\section{Introduction}
\label{problem}
In Direction-of-Arrival (DoA) estimation, coprime arrays offer increased Degrees-of-Freedom (DoF) and enable the identification of more sources than sensors compared to equal-length uniform linear arrays~\cite{LEUS,AMIN3,AMIN5,PP2,PP3,PP4,CHUNLIU3,CHUNLIU5,ZTAN,GOODMAN,PP6,GUO1,GUO2}. Coprime arrays have been successfully employed in applications such as beamforming design~\cite{PP5,CZHOU2, robust_beam} and space-time processing~\cite{CHUNLIU2}, to name a few.
Other non-uniform arrays with increased DoF and closed-form expressions are the nested and MISC arrays~\cite{PP9, zheng2019misc}. 
Intelligent processing of the autocorrelations of the physical array's elements enables the estimation of a signal subspace corresponding to a larger (virtual) array known as ``coarray" which, in general, is non-uniform. Commonly, only a segment of the coarray is retained wherein the elements are uniformly spaced. Alternatively, some works employ interpolation methods to fill the ``gaps'' of the full coarray. In this work, we consider only the uniform segment of the coarray for simplicity. Our result is straightforwardly extended to the full coarray after its gaps are filled by existing interpolation methods (e.g., \cite{FAUZHIA1,CS1,zhou2018coarray}).

A coprime array receiver processes the autocorrelations of the physical-array's elements and estimates the autocorrelation matrix of the coarray. By the received-signal model, the nominal autocorrelation matrix of the coarray has a specific structure: it is (i) Positive Definite (PD), (ii) Hermitian, (iii) Toeplitz, and (iv) its noise-subspace eigenvalues are equal. In practice, the autocorrelations of the physical-array's elements are estimated by processing a collection of received-signal snapshots and diverge from the nominal ones. Accordingly, existing approaches offer autocorrelation-matrix estimates which diverge from the nominal one, while at the same time, violate at least one of the above structure-properties. 

In this work, we propose an optimization framework for computing an improved coarray autocorrelation-matrix estimate that satisfies properties (i)-(iv). In practice, we iteratively solve a sequence of optimization problems, obtaining upon convergence, an improved estimate that satisfies properties (i)-(iv).
Our studies illustrate that the proposed method outperforms standard counterparts, both in estimation error and DoA estimation. 
	
\section{Signal Model And Problem Statement}
\label{signalmodel}
We consider coprime integers~$M<N$ and design coprime array with~$L\eqdef2M+N-1$ elements at locations~$\mathcal L\eqdef\{(i-1)Md\}_{i=1}^N \bigcup\{iNd\}_{i=1}^{2M-1},$ where~$d={\lambda\over 2}$ is the reference inter-element spacing and~$\lambda$ is the wavelength~\cite{PP1}. Narrow-band signals impinge on the array from~${K<MN+M}$ sources with DoAs~$\Theta\eqdef\{\theta_1,\theta_2,\ldots,\theta_K\}$ under carrier frequency~$f_c$ and propagation speed~$c$. Under far-field conditions, the~$k$th source-signal impinges on the array from direction~$\theta_k {\in} (-\frac{\pi}{2}, \frac{\pi}{2}]$ with respect to the broadside. Defining the element-location vector~$\ma p\eqdef \mathrm{sort}(\mathcal{L})$, the array-response vector of source~$k$ becomes~$\mathbf s(\theta_k)\eqdef[  {v(\theta_k)}^{[\ma p]_1}, {v(\theta_k)}^{[\ma p]_2}, \ldots , {v(\theta_k)}^{[\ma p]_L}]^\top,$ where~$v(\theta_k)\eqdef\mathrm{exp}(- \frac{ \imath 2\pi f_c}{c} \sin(\theta_k))$. Accordingly, the receiver collects received-signal snapshots of the form
	\begin{align}
	\centering
	\mathbf y_q \eqdef\sum_{k=1}^{K}\ma s(\theta_k) x_{q,k}+\mathbf n_q \in \mathbb C^{L},
	\label{snapshot}
	\end{align}
	where~$x_{q,k}{ \sim} \mathcal{CN}(0, d_k)$ and~$\mathbf{n}_q {\sim}  \mathcal{CN}(\mathbf 0_L, \sigma^2 \mathbf I_L)$, model the~$q$th symbol transmitted by the~$k$th signal-source (power-scaled and flat-fading-channel processed) and Additive White Gaussian Noise, respectively. Received-symbols are uncorrelated across snapshots and sources. Noise-variables are uncorrelated from received symbols. The receiver's objective is to identify~$\Theta$ from the collected snapshots. Next, we briefly review standard coprime array processing.
  
{\it Physical-Array Autocorrelation Matrix:} The nominal received-signal autocorrelation matrix of the physical array is given by~$\mathbf R_y \eqdef \mathbb E \{\mathbf y_q \mathbf y_q\}=\mathbf S \mathrm{diag}(\mathbf d)\mathbf S^H+\sigma^2\mathbf I_L,$~where~$ \ma S\eqdef[\ma s(\theta_1),\ldots,\ma s(\theta_K)] $ and~$\ma d\eqdef[d_1,\ldots,d_K]^\top$ denote the array-response matrix and source-power vector, respectively. Since~$\Theta$, $\ma d $, and $\sigma^2$ are in practice unknown to the receiver,~$\mathbf R_y$ can not be directly computed and is estimated based on $Q$ received-signal snapshots by~$\widehat{\mathbf R}_y\eqdef\frac{1}{Q}\sum_{q=1}^{Q}\ma y_q\ma y_q^H$.
	
{\it Autocorrelation Sampling:}
Nominally, the receiver processes~$\ma R_y$ and computes the autocorrelation-vector~$\mathbf r\eqdef \mathrm{vec} (\mathbf R_y)=\sum_{k=1}^{K}\ma a(\theta_k) d_k+\sigma ^2  \mathbf i_L,$~where $\mathbf a(\theta_k)\eqdef{\mathbf s(\theta_k)}^* \otimes \mathbf s(\theta_k)$ and $\mathbf i_L \eqdef \mathrm{vec}(\mathbf I_L)$. 
For~$j \in [L^2]\eqdef\{1,\ldots, L^2\}$, it holds~$[\mathbf a(\theta_k)]_j={v(\theta_k)}^{n}$,~$n \in \mathcal{A}\eqdef \mathrm{sort}(\{n_1-n_2\mid n_1,n_2 \in \mathcal{L}\})$. The element-locations of the uniform segment of the coarray are described by
\begin{align}\mathcal{B}\eqdef\{ n\in \mathcal{A} \mid 1-L'\leq n \leq L'-1\},\end{align} where~$ L' \eqdef MN+M$. For every~$n \in \mathcal B$, the receiver discards duplicates by selecting any single index~$j_n \in [L^2]$ such that $[\mathbf a(\theta_k)]_{j_n}={v(\theta_k)}^{n}$. That is, the receiver forms selection-sampling matrix
\begin{align}\mathbf E_{\text{sel}} \eqdef [	\mathbf e_{j_{1-L'}, L^2} , \ldots , \mathbf e_{j_{L'-1}, L^2}],\end{align} where for any~$p \leq P \in \mathbb N_+$,~$\mathbf e_{p,P}$ is the~$p$th column of~$\mathbf I_{P}$, and computes~${\mathbf r}_{\text{sel}} \eqdef  \mathbf E_{\text{sel}}^\top \mathbf r = \sum_{k=1}^K {\mathbf a}_{\text{sel}}(\theta_k) d_k + \sigma^2 \mathbf e_{L', 2L'-1},$ where~${\mathbf a}_{\text{sel}}(\theta_k) \eqdef \mathbf E_{\text{sel}}^\top \mathbf a(\theta_k) = [v(\theta_k)^{1-L'}, \ldots,  v(\theta_k)^{L'-1}] ^\top.$
In practice, the autocorrelation-vector~$\mathbf r$ is estimated by~$  \widehat{\mathbf r}\eqdef \mathrm{vec}(\widehat{  \mathbf R}_y)$ and  ${\ma r}_{\text{sel}}$ is estimated by~$\widehat{{\ma r}}_{\text{sel}}\eqdef\mathbf E_{\text{sel}}^\top\widehat{\mathbf r}$.

{\it Coarray Autocorrelation Matrix:}
	The receiver applies a rank-enhancement approach on $\ma r_{\text{sel}}$ (or, $\widehat{  \mathbf r}_{\text{sel}}$ in practice) to form the autocorrelation matrix of the coarray. Commonly, the \emph{Augmented Matrix} \cite{CHUNLIU1} and \emph{Spatial Smoothing} \cite{PP1} approaches are employed.
	According to the augmented matrix approach, the receiver computes
	\begin{align}
	{\ma R}_{\text{am}} \eqdef \ma F (\ma I_{L'} \otimes {\ma r}_{\text{sel}}) \in \mathbb C^{L' \times L'},
	\label{Z}
	\end{align}
	where
	$\mathbf F \eqdef [\mathbf F_{1}, \mathbf F_{2}, \ldots, \mathbf F_{L'}]$ and, for every $m \in [L']$,  $\mathbf F_m \eqdef [\mathbf 0_{L' \times (L'-m)}, \mathbf I_{L'}, \mathbf 0_{L' \times (m-1)}]$.
	$\ma R_{\text{am}}$ has full-rank, is PD, Hermitian, Toeplitz, and coincides with the autocorrelation matrix of the coarray
	\begin{align}
	\ma R_{\text{co}}= \mathbf S_{\text{co}} \mathrm{diag}(\mathbf d) \mathbf S_{\text{co}}^H + \sigma^2 \mathbf I_{L'},
	\end{align}
	where 
	$[\mathbf S_{\text{co}}]_{m,k} \eqdef v(\theta_k)^{m-1}$, for every $m \in [L']$ and $k \in [K]$. According to the spatial-smoothing approach \cite{PP1}, in the case of known statistics, the receiver computes the spatially-smoothed matrix
	$
	\ma R_{\text{ss}}\eqdef \frac{1}{L'} {\ma R}_{\text{am}}{\ma R}_{\text{am}}^H
	$
which is not an autocorrelation matrix but an autocorrelation matrix is extracted from it as a scaled version of its principal square root \begin{align}
	{\ma R}_{\text{psr}}\eqdef\sqrt{L' }\mathbf R_{\text{ss}}^{\frac{1}{2}}.
	\end{align}
	We notice that ${\ma R}_{\text{am}}{\ma R}_{\text{am}}^H={\ma R}_{\text{am}}^2=L'\ma R_{\text{ss}}$. Moreover, $\ma R_{\text{ss}}$ admits Singular Value Decomposition (SVD) $\ma R_{\text{ss}}\overset{\text{svd}}{=}\ma U \ma \Lambda \mathbf V^H$ which implies that $\ma R_{\text{psr}}=\ma U(\sqrt{L'}\mathbf \Lambda^{\frac{1}{2}})\mathbf V^H={\ma R}_{\text{am}}$. That is, $\ma R_{\text{psr}}$ and ${\ma R}_{\text{am}}$ both coincide with $\Rco$. Here, we note that in the (ideal) case of known statistics to the receiver, all estimates above coincide with the nominal autocorrelation matrix of the coarray and satisfy (i)-(iv). However, in the practical case of unknown statistics (case of interest) to the receiver, the estimates above diverge from~$\Rco$ and satisfy only a subset of (i)-(iv):
The augmented matrix approach of \cite{CHUNLIU1} proposed to substitute the sampling matrix $\mathbf E_{\text{sel}}$ by the averaging sampling matrix 	
\begin{align}
\mathbf E_{\text{avg}} {\eqdef} [	\frac{1}{|\mathcal{J}_{1-L'}|}\sum_{j \in \mathcal{J}_{1-L'}} \mathbf e_{j,L^2}, \ldots , \frac{1}{|\mathcal{J}_{L'-1}|}\sum_{j \in \mathcal{J}_{L'-1}} \mathbf e_{j,L^2}],
\end{align}
where for every $n \in \mathcal B$, $\mathcal{J}_n=\{j \in [L^2]\mid [\ma a(\theta_k)]_j={v(\theta_k)}^n\}$, substituting $\widehat{{\ma r}}_{\text{sel}}$ by $ \widehat{{\ma r}}_\text{avg}\eqdef\mathbf E_{\text{avg}}^\top\widehat{\mathbf r}.$\footnote{\doublespacing If the nominal statistics are known, $\ma r_{\text{sel}}$ and $\ma r_{\text{avg}}=\mathbf E_{\text{avg}}^\top\ma r$ coincide. The latter does not hold if $\ma r_{\text{sel}}$ and $\ma r_{\text{avg}}$ are estimated by $\widehat{\ma r}_{\text{sel}}$ and $\widehat{\ma r}_{\text{avg}}$, respectively.} Accordingly, $\ma R_{\text{co}}$ is estimated by $\widehat{ {\ma R}}_{\text{am}} \eqdef \mathbf F (\mathbf I_{L'} \otimes \widehat{ {\ma r}}_{\text{avg}})$.
	Importantly, it holds that $\widehat{{\ma R}}_{\text{am}}$ is Hermitian and Toeplitz, however, it's not guaranteed to be PD. That is, $\widehat{{\ma R}}_{\text{am}}$ can be an indefinite estimate of $\Rco$ \cite{CHUNLIU1}.
	Similarly, $\ma R_{\text{ss}}$ and $\ma R_{\text{psr}}$ are estimated by
	$\widehat{\ma R}_{\text{ss}}\eqdef \frac{1}{L'} \widetilde{  \mathbf R}_{\text{am}}\widetilde{  \mathbf R}_{\text{am}}^H$ and $\widehat{  \mathbf R}_{\text{psr}}\eqdef \sqrt{L'}\widehat{\ma R}_{\text{ss}}^{\frac{1}{2}}$, respectively, where $\widetilde{\ma R}_{\text{am}}=\ma F(\ma I_{L'}\otimes \widehat{\ma r}_{\text{sel}})$.\footnote{\doublespacing $\widehat{\ma R}_{\text{am}}$ and $\widetilde{\ma R}_{\text{am}}$ denote the augmented matrix approach estimates combined with averaging and selection sampling, respectively.}~$\widetilde{\ma R}_{\text{am}}$ can be an indefinite matrix. In view of the above,  $\widehat{\ma R}_{\text{psr}}$ is by construction a PD and Hermitian matrix estimate of the coarray autocorrelation matrix, however, it violates the Toeplitz structure-property of $\Rco$.
	It follows that $\widehat{\ma R}_{\text{psr}}$ and $\widetilde{\ma R}_{\text{am}}$ no longer coincide, however, their left-hand singular-valued singular vectors span the same signal subspace. For the unknown statistics case, we summarize the above estimates in Table~\ref{tab:mtable}, where for each estimate we mention the employed autocorrelation sampling approach. Moreover, for each structure property guaranteed to be satisfied, we place a \cmark, otherwise, we place a \xmark.  
	Given a coarray autocorrelation matrix estimate $\widehat{\ma R}\in \{\widehat{\ma R}_{\text{am}}, \widetilde{\ma R}_{\text{am}}, \widehat{\ma R}_{\text{psr}}\}$, a standard DoA estimation approach--e.g., MUltiple SIgnal Classification (MUSIC)--is applied for identifying the DoAs in $\Theta$.

\begin{table}[!t]
\centering
\resizebox{1.0\textwidth}{!}{%
\begin{tabular}{@{}llcccc@{}}
\toprule
Matrix estimate                     & Autocorrelation sampling approach & Positive Definite & Toeplitz & Hermitian & Equal noise-subspace eigenvalues \\ \midrule
$\widetilde{\mathbf R}_{\text{am}}$ & Selection                         &           \xmark        &      \cmark    &     \cmark      &           \xmark                       \\
$\widehat{\mathbf R}_{\text{am}}$   & Averaging                         &          \xmark         &      \cmark    &     \cmark      &         \xmark                         \\
$\widehat{\mathbf R}_{\text{psr}}$  & Averaging                         &         \cmark          &       \xmark   &  \cmark          &   \xmark                                \\
 Structured (proposed)               & Averaging                         &         \cmark           &     \cmark     &     \cmark      &        \cmark               
       \\ \bottomrule
\end{tabular}%
}
\caption{Comparison of coarray autocorrelation matrix estimates: autocorrelation sampling approach and structure properties.}
\label{tab:mtable}
\end{table}
	
\section{Proposed Autocorrelation-Matrix Estimate}
\label{proposedwork}
\begin{figure}[!t]
			\centering
		\includegraphics[width=.45\linewidth]{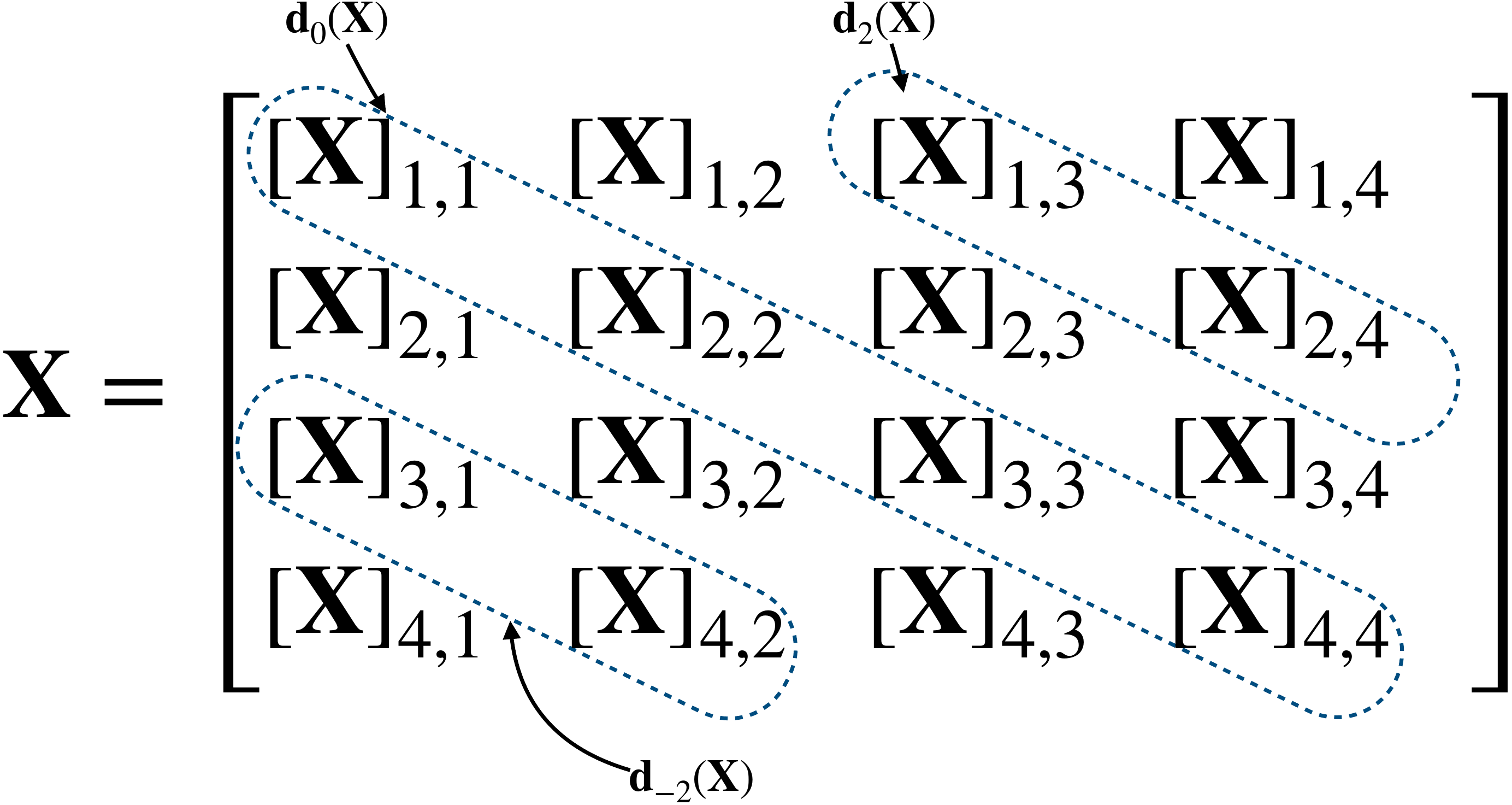}
			\caption{Illustration of the $i$th diagonal of $\mathbf X \in \mathbb C^{4 \times 4}$, $\mathbf d_i(\mathbf X)$, $ i \in \{0,\pm 2\}$.}
			\label{visillust}
		\end{figure}

We propose an algorithm which iteratively solves a sequence of optimization problems returning, upon convegrence, an improved coarray autocorrelation matrix estimate. Motivated by \cite{DGC1}, where it was formally proven that averaging autocorrelation sampling attains superior autocorrelation estimates compared to selection sampling with respect to the MSE metric, we propose to initialize the proposed algorithm to $\ma P_0=\sqrt{\frac{1}{L'}\widehat{\ma R}_{\text{am}}\widehat{\ma R}_{\text{am}}^H}$.
	At iteration  $i\geq 0$, the proposed algorithm computes 
	\begin{align}
	&\ma Q_i=\mPhi{P_i},\label{iteration1}\\
	&\ma R_i=\mPsi{Q_i},\label{iteration2}\\
	&\ma P_{i+1}=\mOmega{ R_i},\label{iteration3}
	\end{align} 
where for any $\ma X=\ma X^H\in \mathbb C^{D \times D}$ with Eigen-Value Decomposition (EVD) $\ma X\overset{\text{evd}}{=}\ma U \mathrm{diag}(\mlambda{X}) \ma U^H$ the following hold.
	\begin{mdef}
		$\mPhi{X}$ returns the nearest Toeplitz matrix, in the Euclidean norm sense\footnote{\doublespacing Otherwise known as Frobenius norm: $\|\cdot\|_F^2$ returns the sum of the squared entries of its argument.}, to $\ma X$: $\mPhi{X}\eqdef\argmin{\ma T_0 \in \mathcal{T}^D}{}{\|\ma X -\ma T_0\|_F^2}$, where
		$\mathcal{T}^D\eqdef\{\mathbf T \in \mathbb{C}^{D \times D}\mid \ma T\text{ is Toeplitz}\}$.
		\label{definition1}
	\end{mdef} 
	\begin{mdef}
		$\mPsi{X}$ returns the nearest Positive Semidefinite (PSD) matrix to $\ma X$: $\mPsi{X}{\eqdef}\argmin{\ma X_0 \in \mathbb{S}^D_+}{\|\ma X -\ma X_0\|_F^2}$, where
		$\mathbb S_{+}^{D}{\eqdef}\{\mathbf A {\in} \mathbb C^{D \times D} {\mid}  \ma A{=}\ma A^H{\succeq }\mathbf 0\}$. 
		\label{definition2}
	\end{mdef} 
	\begin{mdef}
		$\mOmega{X}$ performs an eigenvalue-correction operation to the $D-d$ smallest eigenvalues of $\ma X$. For some general $d \in \{1,\ldots,D-1\}$,
		$
		\mOmega{X}\eqdef\ma U \mathrm{diag}({\mlambdabar{X}})\ma U^H,
		$
		where \begin{align}
		[\mlambdabar{X}]_i=\begin{cases}
		[\mlambda{X}]_i, & i\leq D-d+1, \\
		\frac{1}{D-d}\sum\limits_{i=d+1}^{D}[\mlambda{X}]_i, & i> D-d+1.
		\end{cases}
		\end{align}
		\label{definition3}
	\end{mdef}  
	\noindent   In view of the above, the proposed algorithm seeks to optimize the $D-d$ smallest eigenvalues of the autocorrelation matrix estimate at which it is initialized while preserving the PSD, Hermitian, and Toeplitz structure. Next, we conduct formal convergence analysis of the proposed algorithm. 
	Consider arbitrary $\mathbf X=\ma X^H  \in \mathbb C^{D \times D}$ and let $\mathbf d_i(\mathbf X)$ denote a diagonal of $\mathbf X$ (see Fig. \ref{visillust}) such that 
	\begin{align}
	[\mathbf d_i(\mathbf X)]_j=\begin{cases}
	[\mathbf X]_{j-i,j}, &i \leq 0, \\
	[\mathbf X]_{j,j+i}, &i > 0,
	\end{cases}
	\end{align}
	for any $j \in \{1,2,\ldots,D-|i|\}$.
	The following remarks hold.
	\begin{rem}
		It holds that 
		$
		\mathbf d_i(\mathbf X)=\mathbf d_{i'}^*(\mathbf X), ~\text{if}~ |i|=|i'|.
		$\label{remark1}
	\end{rem}
	\begin{rem}
		Let $\mathbf T\eqdef\mPhi{X}=\min_{\mathbf T_0 \in \mathcal T^D}\|\mathbf X-\mathbf T_0\|_F^2$. For any $i \in \{1-D,\ldots, D-1\}$, it holds that $
		\mathbf d_i(\mathbf T)=\frac{1}{D-|i|}\mathbf 1_{D-|i|}^\top\mathbf d_i(\mathbf X)\mathbf 1_{D-|i|}.
		$
		\label{remark2} 
	\end{rem}
	\begin{rem}
		Let $\mathbf T\eqdef\mPhi{X}$. It holds that 
		$
		\|\mathbf T\|_F^2 \leq \|\mathbf X\|_F^2.
		$
	\end{rem}
	\begin{proof}
	\begin{align}
	\|\mathbf T\|_F^2&=\sum_{i=1-D}^{D-1} \|\mathbf d_i(\mathbf T)\|_2^2
	=\sum_{i=1-D}^{D-1}\|\frac{1}{D-|i|}(\mathbf 1_{D-|i|}^\top\mathbf d_i(\mathbf X))\mathbf 1_{D-|i|}\|_2^2
	=\sum_{i=1-D}^{D-1}\frac{(\mathbf 1_{D-|i|}^\top\mathbf d_i(\mathbf X))^2}{(D-|i|)^2}\|\mathbf 1_{D-|i|}\|_2^2
	\\&=\sum_{i=1-D}^{D-1}\frac{(\mathbf 1_{D-|i|}^\top \mathbf d_i(\mathbf X))^2}{D-|i|}
	\leq \sum_{i=1-D}^{D-1}\frac{\|\mathbf 1_{D-|i|}\|_2^2 \|\mathbf d_i(\mathbf X)\|_2^2}{D-|i|}=\|\mathbf X\|_F^2.
	\end{align}
\end{proof}
	\begin{rem}
		Let $\mathbf T\eqdef\mPhi{X}$ admit EVD\footnote{\doublespacing A Hermitian matrix $\mathbf A$ can be expressed as $\mathbf A{=}\mathbf U\mathbf \Lambda\mathbf U^H$, where $\mathbf \Lambda$ is an upper diagonal with the eigenvalues of $\mathbf A$ in its main diagonal. If $\mathbf A$ is normal (i.e., $\mathbf A\mathbf A^H{=}\mathbf A^H \mathbf A$), then $\mathbf \Lambda$ is diagonal. Every Hermitian matrix is normal.} $\mathbf T=\mathbf U \mathrm{diag}(\boldsymbol \lambda_T)\mathbf U^H$. It holds 
		$
		\|\mathbf T\|_F^2{=}\mathrm{Tr}(\mathbf U\mathrm{diag}(\boldsymbol{\lambda}_T) \mathrm{diag}(\boldsymbol \lambda_T)\mathbf U^H)
		{=}\|\boldsymbol{\lambda}_T\|_2^2.
		$
		\label{remark4}
	\end{rem}
	\begin{rem}
		Let $\mathbf T\eqdef\mPhi{X}$ admit EVD $\mathbf T=\mathbf U \mathrm{diag}(\boldsymbol \lambda_T)\mathbf U^H$ and define $\mathbf P$ such that $
		\mathbf P=\mathbf U \mathrm{diag}(\boldsymbol{\lambda}_P)\mathbf U^H,$
		where $\boldsymbol{\lambda}_P=\boldsymbol{\lambda}_T^+$,--i.e., $\forall i \in [D]$, $[\boldsymbol{\lambda}_P]_i=\max\{[\boldsymbol{\lambda}_T]_i,0\}$. It holds that $\mathbf P$ is the solution to $\underset{\mathbf P_0 \in \mathbb S_+^D}{\text{min.}}\|\mathbf T-\mathbf P_0\|_F^2$.
		A proof for Remark \ref{remark5} was first offered for real matrices in \cite{HIGHAM2}.
		For completeness purposes, we offer an analogous proof for complex-valued matrices.
		\label{remark5}
		\end{rem}
			\begin{proof}
	Consider Hermitian $\ma T \in \mathbb C^{D \times D}$ with EVD $\mathbf T=\mathbf U \mathrm{diag}(\boldsymbol \lambda_T)\mathbf U^H$, $\ma U$ is unitary (i.e., $\ma U\ma U^H=\ma U^H\ma U=\ma I_D$). Let $\ma H=\ma U^H \ma P_0\ma U$ which implies that $\ma P_0=\ma U \ma H \ma U^H$. It holds \begin{align}
	\min_{\mathbf P_0 \in \mathbb S_+^D}\|\mathbf T-\mathbf P_0\|_F^2&=\min_{\ma H \in \mathbb C^{D \times D}}\| \mathrm{diag}(\boldsymbol \lambda_T)-\ma H\|_F^2
	{=}\min_{\ma H \in \mathbb C^{D \times D}} \sum_{\substack{i,j\\i\neq j}} [\ma H]_{i,j}^2{+}\sum_{i=1}^{D}([\boldsymbol \lambda_T]_i{-}[\ma H]_{i,i})^2
	\\&\geq\sum_{\substack{i,j\\i\neq j}} [\ma H]_{i,j}^2{+}\sum_{i=1}^{D}([\boldsymbol \lambda_T]_i{-}[\ma H]_{i,i})
	\geq\sum_{i \in \{1,2,\ldots,D \mid [\boldsymbol{\lambda_T}]_i<0\}}[\boldsymbol{\lambda}_T]_i^2.
	\label{lowerboubd}
	\end{align}
	Similar to \cite{HIGHAM2}, the lower bound in \eqref{lowerboubd} is attained by matrix $\ma H=\mathrm{diag}(\boldsymbol{\lambda}_P)$ for $\boldsymbol{\lambda}_P$ such that $[\boldsymbol{\lambda}_P]_i=\max\{[\boldsymbol{\lambda}_T]_i,0\}$.
	\end{proof}
			\begin{figure}[!t]
		\centering
		\includegraphics[width=.50\linewidth]{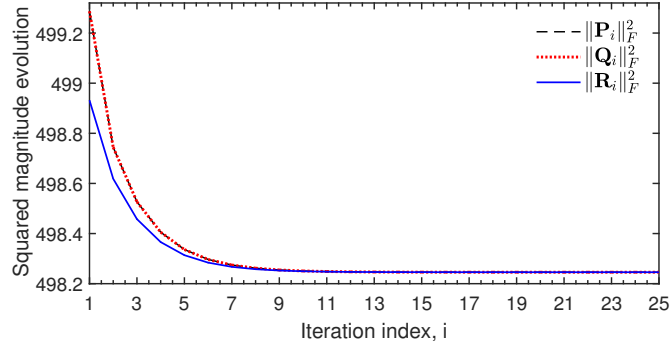}
		\caption{Visual illustration of Proposition \ref{proposition1}. $(M,N){=}(2,3)$, $\Theta=\{-43^\circ,-21^\circ,-10^\circ,17^\circ,29^\circ,54^\circ\}$, $d_k{=}0$ dB $\forall k$, $\sigma^2{=}1$, $Q{=}50$.}
		\label{evolution1}
	\end{figure}%
	\begin{rem}
		For $\ma P=\mPsi{T}$, it holds that 
		$\|\mathbf P\|_F^2\leq\|\mathbf T\|_F^2.$ Formally,
		$
		\|\mathbf P\|_F^2{=}\|\boldsymbol{\lambda}_P\|_2^2=\sum_{i=1}^D[\boldsymbol{\lambda}_P]_i^2\leq \sum_{i=1}^D[\boldsymbol{\lambda}_T]_i^2=\|\mathbf T\|_F^2.
		$
	\end{rem}
	\begin{rem}
		Let $\mathbf A=\mOmega{P}=\mathbf U \mathrm{diag}(\boldsymbol{\lambda}_A)\mathbf U^H,
		$
		where \begin{align}
			\renewcommand{\arraystretch}{1.0}
		[\boldsymbol{\lambda}_A]_i=\begin{cases}
		[\boldsymbol{\lambda}_P]_i, & i\leq D-d+1, \\
		\frac{1}{D-d}\sum\limits_{j=d+1}^{D}[\boldsymbol{\lambda}_P]_i, & i>D-d+1,
		\end{cases}
		\renewcommand{\arraystretch}{1.0}
		\end{align}
		for $d \in \{1,\ldots,D-1\}$. It holds that
		$
		\|\mathbf A\|_F^2\leq\|\mathbf P\|_F^2.
		$
		\label{remark6}
		\end{rem}
		\begin{proof}
				\begin{align}
				\|\mathbf A\|_F^2&{=}\sum_{i=1}^D[\boldsymbol{\lambda}_A]_i^2{=} \sum_{i=1}^{D-d}[\boldsymbol{\lambda}_P]_i^2+d(\frac{1}{d}\sum_{j=D-d+1}^{D}[\boldsymbol{\lambda}_P]_j)^2
				\leq \sum_{i=1}^{D-d}[\boldsymbol{\lambda}_P]_i^2+\sum_{j=D-d+1}^{D}[\boldsymbol{\lambda}_P]_j^2=\|\mathbf P\|_F^2.
				\end{align}
		\end{proof}
		\noindent In view of Remarks \ref{remark1}-\ref{remark6}, the following Proposition derives. 
\begin{proposition}
For $\ma Q_i, \ma R_i$, and $\ma P_{i+1}$ in \eqref{iteration1}-\eqref{iteration3}, it holds $\|\ma Q_i\|_F^2{\geq} \|\ma R_i\|_F^2{\geq} \|\ma P_{i+1}\|_F^2{\geq}\|\ma Q_{i+1}\|_F^2\geq\ldots\geq 0 \forall i{\geq} 0.$
		\label{proposition1}
	\end{proposition} 
		
\noindent Proposition \ref{proposition1} states that the proposed algorithm is guaranteed to converge. In practice, one can terminate the iterations when $\|\ma P_{i+1}-\ma P_{i}\|\leq \epsilon$, for some $\epsilon\geq0$. For sufficiently small $\epsilon$, Proposition 1 implies that, at convergence, $\ma P_{i+1}=\ma R_{i}=\ma Q_{i}$ which, in turn, implies that the algorithm converged to a PD, Hermitian, and Toeplitz matrix the noise-subspace eigenvalues of which  are equal. A visual illustration of Proposition \ref{proposition1} and a pseudocode of the proposed algorithm are offered in Fig. \ref{evolution1} and  Fig. \ref{algo}, respectively. Importantly, $\forall i\geq 0$, the Algorithm of Fig. \ref{algo} computes $\ma Q_i, \ma R_i, \ma P_{i+1}$ by closed-form expressions with cost at most the cost of EVD--i.e., $\mathcal O(D^3)$. Overall, the cost of the proposed algorithm is $\mathcal O(TD^3)$ where $T$  is the number of iterations required for convergence.
	\begin{figure}[t!]
		\begin{flushleft}
			\renewcommand{\arraystretch}{1.0}
			{\small
				{\hrule height 0.2mm} 
				\vspace{0.03cm}
				{\hrule height 0.2mm}
				\vspace{0.025cm}
				{\bf{Algorithm 1.} \normalfont{Structured coarray autocorrelation matrix estimation}}
				{\hrule height 0.2mm}
				\vspace{0.03cm}
				{\hrule height 0.2mm}
				\textbf{Input:} Coarray autocorrelation matrix estimate $ \widehat{\ma R} $ \\
				\begin{tabular}{r l l}
					0: & $\ma P_0\leftarrow \widehat{\ma R}$~~~~~~~~~~~\hspace{1 pt}\% \texttt{\footnotesize{Initialization}}\\
					1: & Until convergence/termination \\
					2: & ~~~~ $\ma Q_i\leftarrow\mPhi{ P_i}$  ~~\hspace{3 pt}\% \texttt{\footnotesize{Nearest Toeplitz to }}$\ma P_i$\\    
					3: & ~~~~ $\ma R_i\leftarrow\mPsi{ Q_i}$  ~~\hspace{2 pt}\% \texttt{\footnotesize{Nearest PSD to} }$\ma Q_i$\\
					4: & ~~~~ $\ma P_{i+1}\leftarrow\mOmega{ R_i}$ ~~\hspace{-5 pt}\% \texttt{\footnotesize{Eigenvalue-correction}}\\  
				\end{tabular} \\
				\textbf{Return:} $\ma R_{\text{}}\leftarrow \ma P_i$
				\vspace{0.5mm}
				{\hrule height 0.2mm} 
				\vspace{0.03cm}
				{\hrule height 0.2mm}  		
			} 
			\vspace{0.45cm}
			\caption{\small Proposed structured coarray autocorrelation matrix estimation.}
			\renewcommand{\arraystretch}{1}
			\label{algo}
		    \end{flushleft}
		\end{figure}%
	\section{Numerical Studies}	
	\label{numericalstudies}
\begin{figure}[!t]
			\begin{subfigure}{.32\textwidth}
				\includegraphics[width=1.0\textwidth]{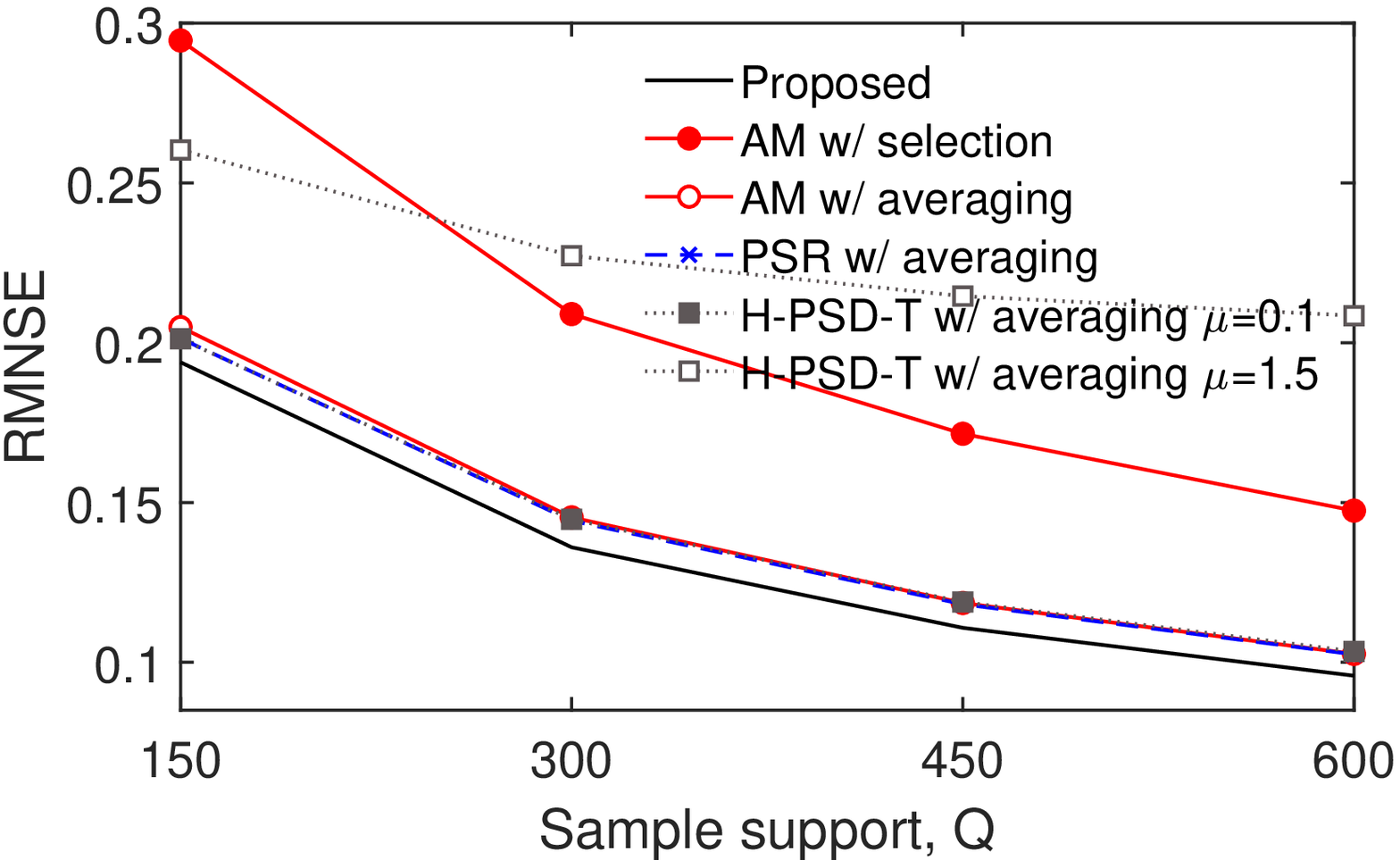}
				\caption{$\mathrm{SNR}=-4$dB.}
				\label{RMNSEm4dB}
			\end{subfigure}
			\begin{subfigure}{.32\textwidth}
				\includegraphics[width=1.0\textwidth]{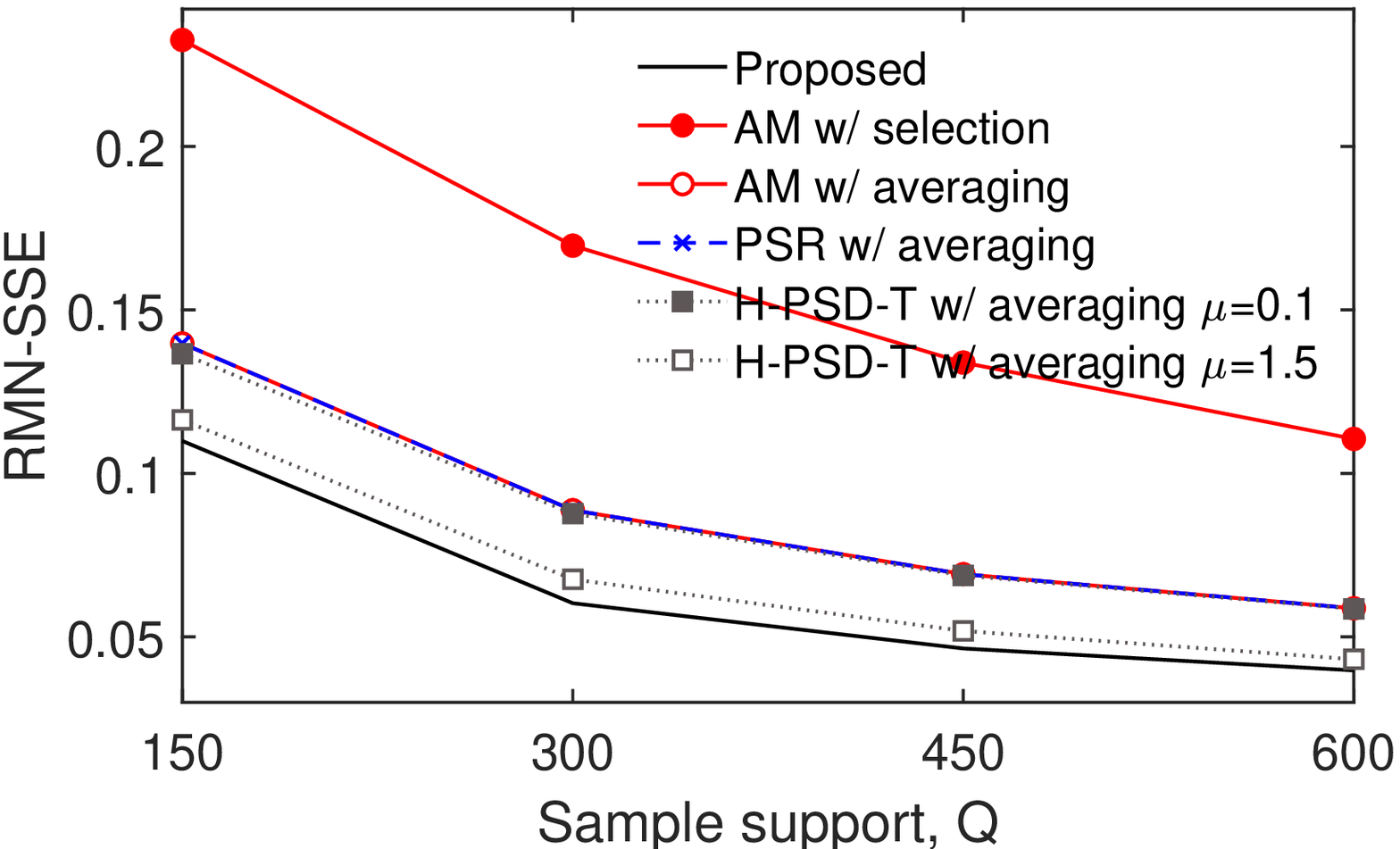}
				\caption{$\mathrm{SNR}=-4$dB.}
				\label{RMNSSEm4dB}
			\end{subfigure}
			\begin{subfigure}{.32\textwidth}
				\includegraphics[width=1.0\textwidth]{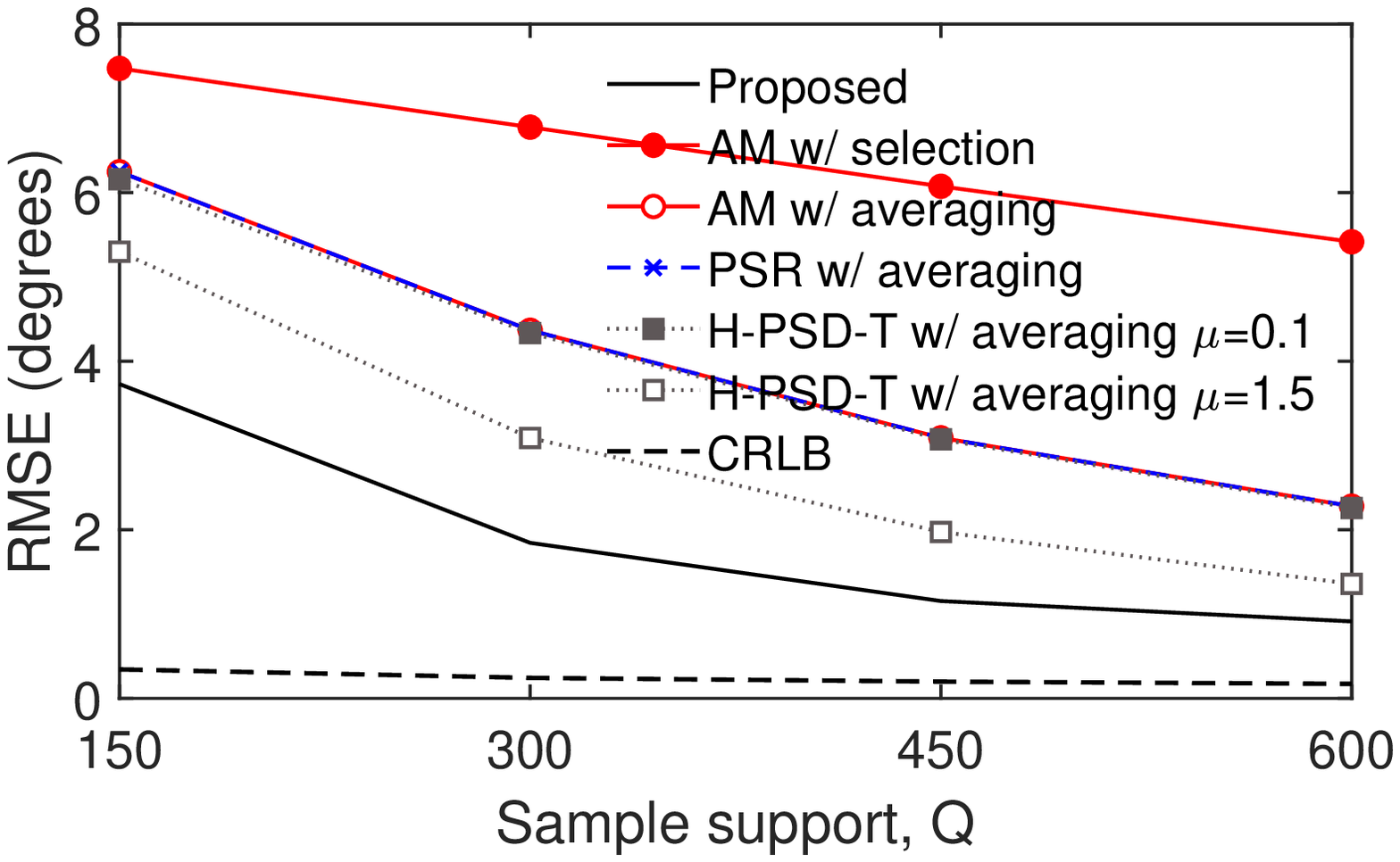}
				\caption{$\mathrm{SNR}=-4$dB.}
				\label{RMSEm4dB}
			\end{subfigure}
			\begin{subfigure}{.32\textwidth}
				\includegraphics[width=1.0\textwidth]{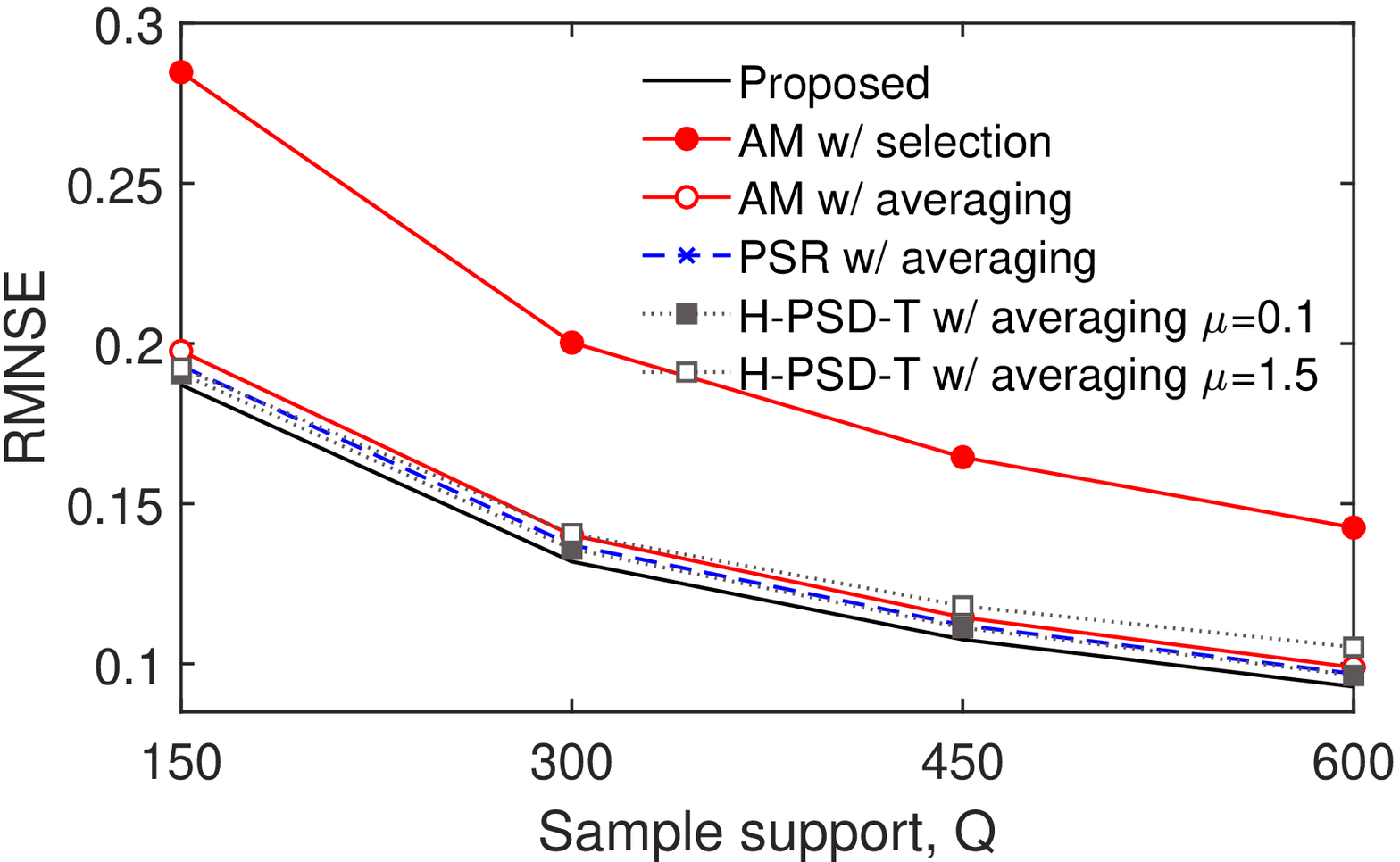}
				\caption{$\mathrm{SNR}=2$dB.}
				\label{RMNSE2dB}
			\end{subfigure}\quad
			\begin{subfigure}{.32\textwidth}
				\includegraphics[width=1.0\textwidth]{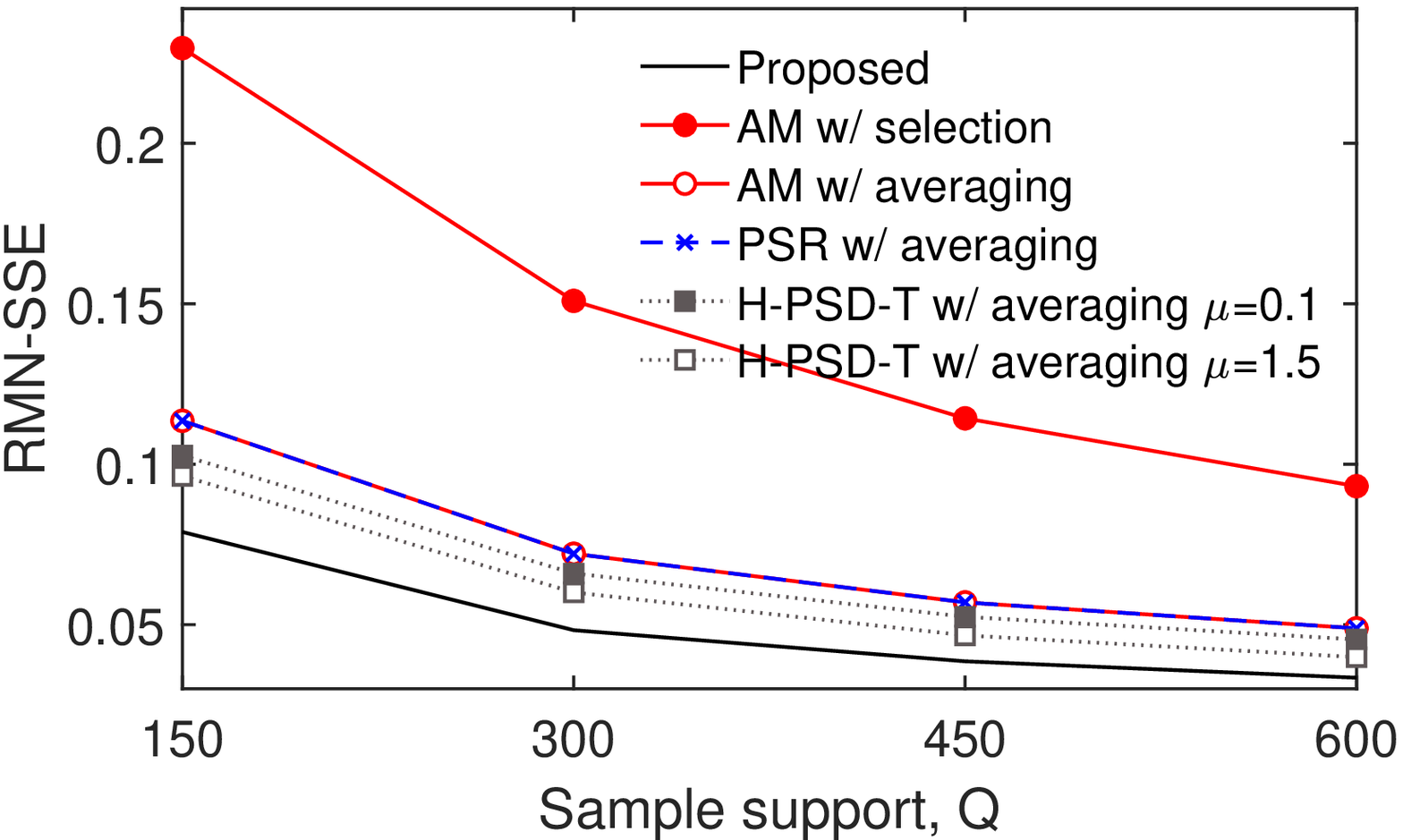}
				\caption{$\mathrm{SNR}=2$dB.}
				\label{RMNSSE2dB}
			\end{subfigure}\quad
			\begin{subfigure}{.32\textwidth}
				\includegraphics[width=1.0\textwidth]{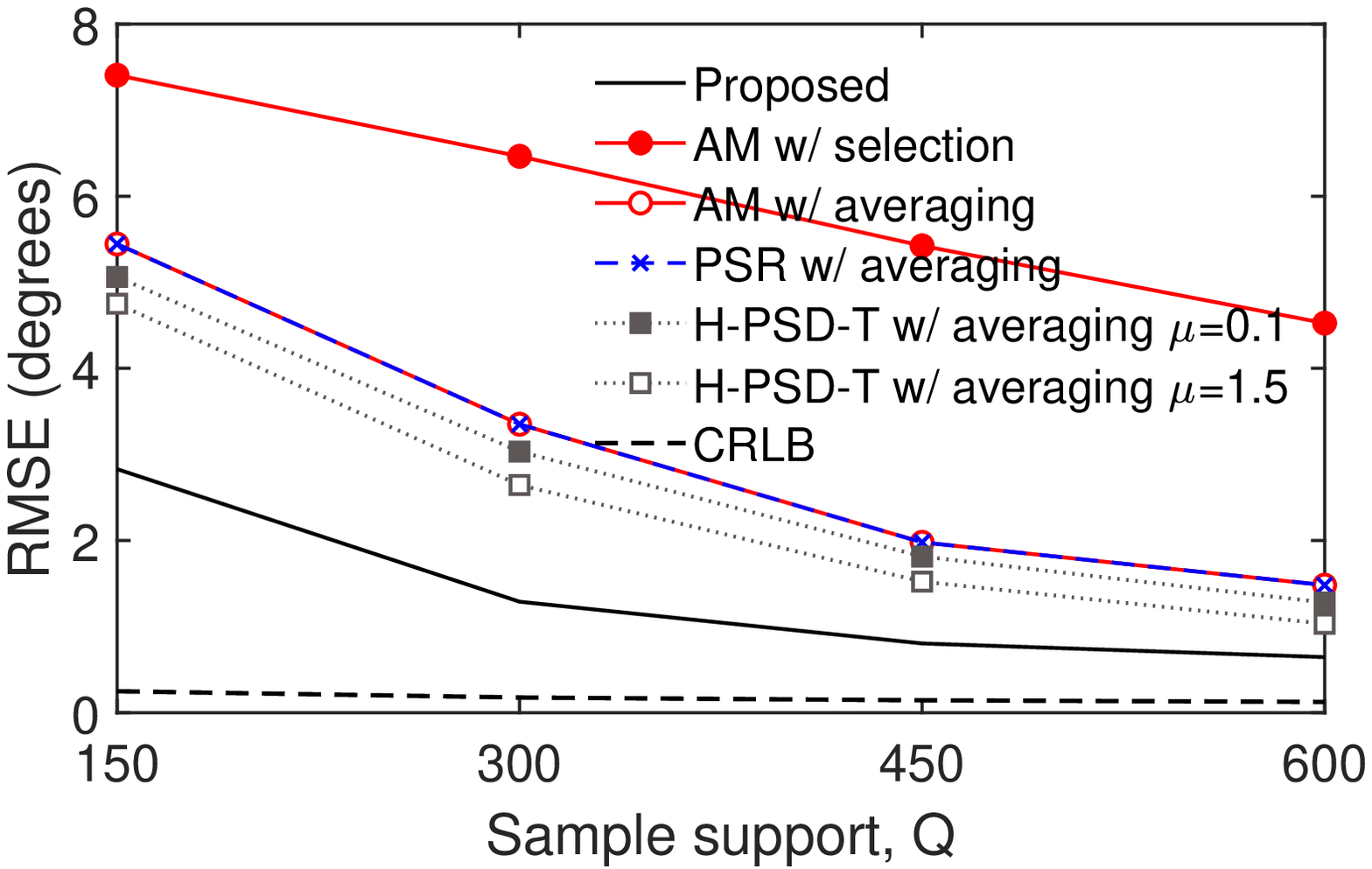} 
				\caption{$\mathrm{SNR}=2$dB.}
				\label{RMSE2dB}
			\end{subfigure}%
			\caption{RMNSE, RMN-SSE, and RMSE metrics versus sample support, $Q$ for $\mathrm{SNR} \in \{-4,2\}$dB.}
		\end{figure}

We consider coprime naturals $(M,N)=(3,5)$ and form coprime array with $L=10$ elements. Source-signals impinge on the array from $K=13$ DoAs $\{\theta_k\}_{k=1}^{13}$,  $\theta_k=(-75+(k-1)12)^\circ$. The noise variance is set to $\sigma^2=0$dB. All sources emit signals with equal power $d_k=\alpha^2$dB. Accordingly, the Signal-to-Noise Ratio $\mathrm{SNR}= \alpha^2$. The receiver collects $Q \in \{150,300,450,600\}$ received-signal snapshots. For every $Q$, we consider 3000 statistically independent realizations of noise; i.e., $\{\ma y_{r,1},\ldots,\ma y_{r,Q}\}_{r=1}^{3000}$. At every realization $r$, we compute coarray autocorrelation matrix estimates corresponding to the augmented matrix approach (AM), principal square root of the spatial smoothed matrix (PSR), nearest Hermitian, PSD, and Toeplitz (H-PSD-T) approach of \cite{zhou2018coarray}\footnote{\doublespacing H-PSD-T seeks a Hermitian-PSD-Toeplitz matrix which fills the gaps of the coarray. When the uniform segment of the coarray is considered, H-PSD-T returns $\argmin{\ma R \in \mathbb S_+^{L'}}{\|\ma R-\widehat{\ma R}_{\text{am}}\|_F^2{+}\mu \|\ma R\|_*}$, where $\mu\|\ma R\|_*$ is a regularization term that moderates overfitting.}, and the proposed structured estimate.  For every method and estimate $\widehat{  \mathbf R}_r$ at realization $r$, we compute the normalized squared error $\mathrm{NSE}(\widehat{  \mathbf R}_r)=\|\widehat{  \mathbf R}_r-\Rco\|_F^2 \|\Rco\|_F^{-2}$. Then, we compute the Root Mean Normalized Squared Error  $\text{RMNSE}=\sqrt{\frac{1}{3000}\sum_{r=1}^{3000}\mathrm{NSE}(\widehat{  \mathbf R}_r)}$. In Fig. \ref{RMNSEm4dB} and Fig. \ref{RMNSE2dB}, we plot the  RMNSE versus sample support, $Q$, for $\mathrm{SNR}=-4$dB and $\mathrm{SNR}=2$dB, respectively. Expectedly, we observe that all methods employing averaging-sampling perform similarly well. The proposed estimate attains superior estimation performance across the board. Moreover, we notice the sensitivity of H-PSD-T with respect to the ad-hoc parameter $\mu$; e.g., for $\mathrm{SNR}=-4$dB, H-PSD-T with $\mu=1.5$ exhibits low performance while for  $\mathrm{SNR}=2$dB it exhibits high estimation performance.
	
Thereafter, we consider that the nominal coarray autocorrelation matrix admits SVD $\Rco=\ma Q_{\text{co}}\ma \Sigma_{\text{co}}\ma V_{\text{co}}^H+\sigma^2\bar{\ma Q}_{\text{co}}\bar{\ma Q}_{\text{co}}^H$, where $\ma Q_{\text{co}}$ and $\bar{\ma Q}_{\text{co}}$ correspond to the signal and noise subspace bases, respectively. Similarly, every coarray autocorrelation matrix estimate $\widehat{  \mathbf R}_r$ admits SVD $\widehat{  \mathbf R}_r=\ma Q_r\ma \Sigma_r \ma V^H_r +\bar{\ma Q}_r\bar{\ma \Sigma}_r\bar{\ma V}_r^H$, where $\ma Q_r$ denotes the signal-subspace-basis of the $K$ dominant left-hand singular valued singular vectors of $\widehat{  \mathbf R}_r$. At each realization and for every value of $Q$, we compute the normalized squared subspace error  $\mathrm{NSSE}(\widehat{  \mathbf Q}_r)=\|\widehat{  \mathbf Q}_r\widehat{  \mathbf Q}_r^H-\mathbf Q_{\text{co}}\mathbf Q_{\text{co}}^H\|_F^2 (2K)^{-1}$. Then, we compute the Root Mean Normalized Squared Subspace Error  $\text{RMN-SSE}=\sqrt{\frac{1}{3000}\sum_{r=1}^{3000}\mathrm{NSSE}(\widehat{  \mathbf Q}_r)}$. In Fig. \ref{RMNSSEm4dB} and Fig. \ref{RMNSSE2dB}, we plot the RMN-SSE versus sample support for $\mathrm{SNR}=-4$dB and $2$dB, respectively. We notice the influence of the ad-hoc parameter $\mu$ with respect to H-PSD-T and observe that the proposed structured estimate clearly outperforms all counterparts across the board in subspace estimation performance. 
Next, for every value of sample support and realization $r$, we conduct DoA estimation by applying MUSIC on the estimate $\widehat{  \mathbf R}_r$ which returns estimates $\{\widehat{\theta}_{k,r}\}_{k=1}^{13}$~\cite{DGC1}. Then, we measure the Root Mean-Squared-Error $\mathrm{RMSE}=\sqrt{\frac{1}{3000}\frac{1}{13} \sum_{r=1}^{3000}\sum_{k=1}^{13} (\theta_k-\widehat{  \theta}_{k,r})^2}$ and illustrate the corresponding RMSE curves versus sample support $Q$, in Fig. \ref{RMSEm4dB} and Fig. \ref{RMSE2dB}, for $\mathrm{SNR}= -4$dB and $2$dB, respectively. We include the Cram\'er Rao Lower Bound (CRLB) curves as benchmarks \cite{MWANG}. We notice that the the performances of standard counterparts (AM, PSR) deviate significantly from the CRLB. In contrast, the proposed coarray autocorrelation matrix estimate outperforms all counterparts by at least $0.3^\circ$ and at most $2^\circ$. In addition, as $Q$ increases, its performance curves approach the CRLB curves.
\section{Conclusions}
\label{conclusions}
We proposed an optimization framework which computes a structured coarray autocorrelation matrix estimate. The proposed algorithm is accompanied by convergence analysis and is guaranteed to return a a coarray autocorrelation matrix estimate satisfying all structure properties of the true autocorrelation matrix. Numerical studies illustrate the enhanced performance of the proposed estimate compared to standard counterparts, both in autocorrelation matrix estimation error and DoA estimation. 

\bibliographystyle{IEEEtran}
\bibliography{structured_coprime_arXiv}
\flushend
\end{document}